\documentclass[english]{revtex4-2}
\usepackage[T1]{fontenc}
\usepackage[latin9]{inputenc}
\usepackage{babel}
\usepackage{float}
\usepackage{units}
\usepackage{textcomp}
\usepackage{amsmath}
\usepackage{graphicx}
\usepackage[unicode=true,pdfusetitle,
 bookmarks=true,bookmarksnumbered=false,bookmarksopen=false,
 breaklinks=false,pdfborder={0 0 1},backref=false,colorlinks=false]
 {hyperref}

\makeatletter
\usepackage{slashed}

\@ifundefined{showcaptionsetup}{}{%
 \PassOptionsToPackage{caption=false}{subfig}}
\usepackage{subfig}
\makeatother

\begin{document}
\title{Renormalization group improvement of the effective potential in a
$\left(1+1\right)$ dimensional Gross-Neveu model}
\author{A. G. Quinto}
\email{aquinto@uninorte.edu.co}

\affiliation{Departamento de Física y Geociencias, Universidad del Norte, Km. 5
Vía Antigua Puerto Colombia, Barranquilla 080020, Colombia}
\affiliation{Facultad de Ciencias Básicas, Universidad del Atlántico Km. 7, Via
a Pto. Colombia, Barranquilla, Colombia}
\author{R. Vega Monroy}
\email{ricardovega@mail.uniatlantico.edu.co}

\affiliation{Facultad de Ciencias Básicas, Universidad del Atlántico Km. 7, Via
a Pto. Colombia, Barranquilla, Colombia}
\author{A. F. Ferrari}
\email{alysson.ferrari@ufabc.edu.br}

\affiliation{Centro de Ciências Naturais e Humanas, Universidade Federal do ABC--UFABC,
Rua Santa Adélia, 166, 09210-170, Santo André, SP, Brazil}
\begin{abstract}
In this work, we investigate the consequences of the Renormalization
Group Equation (RGE) in the determination of the effective potential
and the study of Dynamical Symmetry Breaking (DSB) in an Gross-Neveu
(GN) model with $N$ fermions fields in $(1+1)$ dimensional space-time,
which can be applied as a model to describe certain properties of
the polyacetylene. The classical Lagrangian of the model is scale
invariant, but radiative corrections to the effective potential can
lead to dimensional transmutation, when a dimensionless parameter
(coupling constant) of the classical Lagrangian is exchanged for a
dimensionful one, a dynamically generated mass for the fermion fields.
For the model we are considering, perturbative calculations of the
effective potential and renormalization group functions up to three
loops are available, but we use the RGE and the leading logs approximation
to calculate an improved effective potential, including contributions
up to six loops orders. We then perform a systematic study of the
general aspects of DSB in the GN model with finite N, comparing the
results we obtain with the ones derived from the original unimproved
effective potential we started with.
\end{abstract}
\maketitle

\section{Introduction}

In quantum field theory Dynamical Symmetry Breaking (DSB) is a key
mechanism that has applications in particle physics and condensed
matter systems\,\citep{Gross,Chodos1994,Campbell1982}, where quantum
corrections are entirely responsible for the appearance of nontrivial
minima of the effective potential. In the case of particle physics,
for example, we have a Higgs mechanism playing a fundamental role
in the Standard Model: in this case, the symmetry breaking requires
a mass parameter in the tree-level Lagrangian, but Coleman and Weinberg
(CW) demonstrated\,\citep{Coleman1973} that spontaneous symmetry
breaking may occur due to radiative corrections even when this mass
parameter is absent from the Lagrangian (which is, therefore, scale
invariant). For the study of the CW mechanism, we need to calculate
the effective potential, a powerful tool to explore many aspects of
the low-energy sector of a quantum field theory. In many cases, the
one-loop approximation is good enough, but it can be improved it,
by adding higher order contributions in the loop expansion. A standard
tool for improving a perturbative calculation performed up to some
loop level is the Renormalization Group Equation (RGE) which, together
with a reorganization of the perturbative results in terms of leading
logs, have been shown to be very effective\,\citep{AHMADY2003221,PhysRevD.72.037902,Souza2020,Quinto2016,Dias2014,CHISHTIE2007}.
We refer the reader to section 3 in\,\citep{Quinto2016} for a short
review of the method, and\,\citep{Elias2003,Chishtie2006,Chishtie2011,Steele2013}
for some of the interesting results that have been reported with the
use of the RG improvement.

The Gross-Neveu (GN) model with $N=2$ fermions has great relevance
in the study of the polyacetylene, $\left(CH\right)_{x}$, which is
a polymer with conductive properties which are acquired through doping\,\citep{Chiang1977}.
Polyacetylene is a straight chain that can have two forms, trans and
cis. The trans form (trans-polyacetylene), which is the most stable,
has a doubly degenerate ground state. These circumstances allow the
existence of topological excitations, which entails a great phenomenological
richness in this type of models. In\,\citep{Campbell1982,Takayama1980}
it was shown that in the continuous limit, and in the approximation
where the dynamical vibration of lattice (phonons) is ignored, the
metal-insulator transition in the polyacetylene can be described by
the GN model in $N=2$. In addition, the polyacetylene exhibits some
remarkable effects, such as the Peierls mechanism\,\citep{Horovitz1993},
which is the generation of an energy gap for electrons through the
coupling with phonons. This mechanism is analogous to the Yukawa interaction
in the Standard Model.

The GN model can be seen as an effective low energy theory for the
polyacetylene. This was shown by the Takayama-Lin-Liu-Maki (TLM) model\,\citep{Takayama1980},
where the effective low-energy theories of the Su-Schrieffer-Heeger
(SSH) model\,\citep{Su1979} are described by a theory of four fermions
fields in $\left(1+1\right)$ dimensions. In this model the behavior
of the energy band (gap) $\Delta$ is described as,

\begin{align}
\Delta & =W\exp\left(-\pi v_{f}w_{Q}^{2}/g_{TLM}^{2}\right),\label{eq:Gap nergy}
\end{align}
where $W$ is the width of the energy band, $v_{f}$ is the Fermi
velocity, $w_{g}^{2}/g_{TLM}^{2}$ is the coupling constant between
the electron and phonons. If the adiabatic approximation is used in
the TLM model, then it can be related to the GN model, and therefore,
we can find an expression analogous to Eq.\,(\ref{eq:Gap nergy}),
which is related to the mass obtained in GN by symmetry breaking,
\begin{align}
m & =g_{GN}\sigma_{0}=2\Lambda\exp\left(-\pi/Ng_{GN}^{2}\right),\label{eq:mass in GN model}
\end{align}
$\sigma_{0}$ being a constant scalar field and $\Lambda$ a renormalization
scale, which is not a physical parameter, and therefore the only quantity
that is measured is the mass, $m$. On the other hand, if we compare
this with its analog, Eq.\,(\ref{eq:mass in GN model}), $\Delta$
and $W$ are parameters measured in $\left(CH\right)_{x}$. Now comparing
(\ref{eq:Gap nergy}) and (\ref{eq:mass in GN model}) we can observe
a relationship between the coupling constants,
\begin{align}
g_{GN}^{2} & =\frac{g_{TLM}^{2}}{2v_{f}w_{Q}^{2}},\label{eq:g-NG}
\end{align}
where we have replaced $N=2$, which is the relevant value for the
description of the polyacetylene.

Our goal here is to study, via radiative corrections, the generation
of mass by DSB. In this case the mass will be obtained by

\begin{equation}
m_{\sigma}^{2}=\left.\frac{d^{2}}{d\sigma^{2}}V_{\mathrm{eff}}\left(\sigma\right)\right|_{\sigma=\mu},\label{eq:massa}
\end{equation}
where $\mu$ is the renormalization scale introduced in our model
by regularization, and $V_{\mathrm{eff}}\left(\sigma\right)$ is the
effective potential which is a function of the (classical) scalar
field $\sigma$.

In this paper, we considered the three loops calculation of renormalization
group functions and effective potential for the $\left(1+1\right)$
dimensional GN model with finite $N$ (that is to say, without recourse
to the $1/N$ expansion) that have been described in\,\citep{Luperini1991}.
The RGE is then used to improve this calculation, incorporating terms
that originate from higher loop orders (up to six). Then, we study
the DSB properties of the model using the unimproved (directly obtained
by perturbative calculations) and RGE improved effective potentials,
and we observe that the improvement of the effective potential leads
to relevant differences in comparison with the unimproved one found
in the literature.

There have been many studies of the $\left(1+1\right)$ GN model in
the literature, usually considering the $1/N$ expansion. In this
regard, Ref.\,\citep{Blaizot:2002nh} presents a nice review of leading
and sub-leading orders in this expansion, at finite temperature. The
phase diagram for the model has been first established in Ref.\,\citep{Wolff:1985av},
and recently revised by lattice computations\,\citep{Pannullo:2019bfn,Pannullo:2019prx,Lenz:2020bxk,Lenz:2020cuv}.
Another recent study of this phase diagram, using mean-field techniques,
is presented in\,\citep{Stoll:2021ori}. Finally, studies using the
functional renormalization group have also been reported\,\citep{Stoll:2021ori}.
Our approach is complementary, for not resorting to the $1/N$ expansion,
thus being particularly adequate for models with small $N$; on the
other hand, it is inherently perturbative. It is also interesting
to notice that we work at zero temperature and chemical potential,
so we investigate a single point in the phase diagram that was discussed
in the above-mentioned works. But, at this single point, we are able
to perform calculations analytically. Generalizations for finite temperature
and chemical potential are possible, but not trivial, and are left
for future works.

This paper is organized as follows: in section.\,\ref{sec:2} we
present our model with the renormalization group functions and unimproved
effective potential found in the literature. In section\,\ref{sec:Improvement-of-effective}
we calculate the improvement of effective potential using the standard
approach of RGE, and section\,\ref{sec:Dynamical-symmetric-breaking}
is devoted to study DSB in our model. In section\,\ref{sec:Conclusion}
we present our conclusions and future perspectives.

\section{\label{sec:2}Renormalization group functions and unimproved effective
potential for Gross-Neveu model}

We start with the Euclidean formulation of the massless $\left(1+1\right)$
dimensional GN model studied by Luperini and Rossi\,\,\citep{Luperini1991}
whose Lagrangian with $N$ fermions fields and $U\left(N\right)$
symmetry is, 
\begin{align}
\mathcal{L}_{1} & =\overline{\psi}\slashed{\partial}\psi-\frac{1}{2}g\left(\overline{\psi}\psi\right)^{2}.\label{eq:GN model}
\end{align}
This model has a discrete $\text{\ensuremath{\gamma_{5}}}$ invariance
$\psi\rightarrow\exp\left[\left(\nicefrac{\pi}{2}\right)\gamma_{5}\right]\psi$,
whose spontaneous breakdown leads to a nonzero vacuum expectation
value $\left\langle \bar{\psi}\psi\right\rangle $ and thus to a dynamical
mass generation\,\citep{Gross}. Also, the model is known to be asymptotically
free in two dimensions, and can be extended to, 
\begin{align}
\mathcal{L}_{2} & =\bar{\psi}\slashed{\partial}\psi+\sigma\bar{\psi}\psi+\frac{1}{2}\frac{\sigma^{2}}{g}-\frac{1}{2}h\left(\bar{\psi}\psi\right)^{2},\label{eq:GN modified model}
\end{align}
where $\sigma$ is the scalar field, $\psi$ is the fermion field,
$g$ and $h$ are dimensionless coupling constants that appear with
the introduction of the auxiliary field $\sigma$, which carries the
same quantum numbers as $\bar{\psi}\psi$, i.e. $\sigma=-g\bar{\psi}\psi$.
The Lagrangians $\mathcal{L}_{1}$ and $\mathcal{L}_{2}$ are equivalent
both at classical level (using the equations of motion for $\sigma$
in\,(\ref{eq:GN modified model}) to obtain\,(\ref{eq:GN model}))
as well as at the quantum level, since a gaussian integration over
$\sigma$ in the partition function calculated from \,(\ref{eq:GN modified model})
leads to the same partition function derived by\,\,(\ref{eq:GN model}).

The renormalization group functions $\beta$ and $\gamma$ were calculated
for this model up to three loop order (see the Ref.\,\citep{Luperini1991}
for more details), and we quote the result,
\begin{align}
\beta_{g}\left(g,h\right) & =\beta_{g}^{\left(2\right)}\left(g,h\right)+\beta_{g}^{\left(3\right)}\left(g,h\right)+\beta_{g}^{\left(4\right)}\left(g,h\right),\label{eq:Beta g}
\end{align}
where\begin{subequations} 
\begin{align}
\beta_{g}^{\left(2\right)}\left(g,h\right) & =\left(1-2N\right)gh+\left(1-N\right)g^{2},\\
\beta_{g}^{\left(3\right)}\left(g,h\right) & =\left(N-\frac{1}{2}\right)g^{2}h+\frac{1}{4}\left(2N-1\right)g^{3}+\frac{1}{4}\left(2N-1\right)gh^{2},\\
\beta_{g}^{\left(4\right)}\left(g,h\right) & =\frac{1}{16}\left(3-7N+2N^{2}\right)g^{4}+\frac{9}{16}\left(1-3N+2N^{2}\right)g^{2}h^{2}\nonumber \\
 & +\frac{3}{16}\left(3-8N+4N^{2}\right)g^{3}h+\frac{1}{16}\left(3-10N+8N^{2}\right)gh^{3},
\end{align}
\end{subequations}

\begin{align}
\beta_{h}\left(g,h\right) & =\beta_{h}^{\left(2\right)}\left(g,h\right)+\beta_{h}^{\left(3\right)}\left(g,h\right)+\beta_{h}^{\left(4\right)}\left(g,h\right),\label{eq:Beta h}
\end{align}
where\begin{subequations} 
\begin{align}
\beta_{h}^{\left(2\right)}\left(g,h\right) & =gh+\left(1-N\right)h^{2},\\
\beta_{h}^{\left(3\right)}\left(g,h\right) & =-\frac{1}{4}\,g^{3}+\frac{1}{2}\left(N-2\right)g^{2}h+\left(N-\frac{5}{4}\right)gh^{2}-\frac{1}{2}\left(1-N\right)h^{3},\\
\beta_{h}^{\left(4\right)}\left(g,h\right) & =\frac{1}{16}\left(25-26N\right)gh^{3}+\frac{1}{8}\left(2-N\right)g^{4}+\frac{1}{16}\left(19-12N-4N^{2}\right)g^{3}h\nonumber \\
 & +\frac{1}{16}\left(7-9N+2N^{2}\right)h^{4}-\frac{3}{16}\left(-11+9N+2N^{2}\right)g^{2}h^{2},
\end{align}
\end{subequations}

\begin{align}
\gamma_{\sigma}\left(g,h\right) & =\gamma_{\sigma}^{\left(1\right)}\left(g,h\right)+\gamma_{\sigma}^{\left(2\right)}\left(g,h\right)+\gamma_{\sigma}^{\left(3\right)}\left(g,h\right),\label{eq:Gamma sigma}
\end{align}
where\begin{subequations}

\begin{align}
\gamma_{\sigma}^{\left(1\right)}\left(g,h\right) & =-\frac{1}{2}\,g+\left(N-\frac{1}{2}\right)h,\\
\gamma_{\sigma}^{\left(2\right)}\left(g,h\right) & =\frac{1}{8}\left(1-2N\right)g^{2}+\frac{1}{4}\left(1-2N\right)gh+\frac{1}{8}\left(1-2N\right)h^{2},\\
\gamma_{\sigma}^{\left(3\right)}\left(g,h\right) & =\frac{9}{32}\left(2N-1\right)g^{2}h+\frac{1}{32}\left(-3+10N-8N^{2}\right)h^{3}\nonumber \\
 & -\frac{3}{32}\left(3-8N+4N^{2}\right)gh^{2}+\frac{1}{32}\left(-3+4N+4N^{2}\right)g^{3},
\end{align}
\end{subequations} and

\begin{align}
\gamma_{R}\left(g,h\right) & =\gamma_{R}^{\left(1\right)}\left(g,h\right)+\gamma_{R}^{\left(3\right)}\left(g,h\right),\label{eq:Gamma R}
\end{align}
with\begin{subequations} 
\begin{align}
\gamma_{R}^{\left(1\right)}\left(g,h\right)= & N\,g,\\
\gamma_{R}^{\left(3\right)}\left(g,h\right)= & \frac{3}{16}\left(1-2N\right)Ng^{3}+\frac{3}{8}\left(1-2N\right)N\,g^{2}h+\frac{3}{16}\left(1-2N\right)N\,gh^{2}.
\end{align}
\end{subequations}

In the previous equations, the superscript mean the global power of
coupling constant in each term. This notation will be usefull to organize
the terms for the calculation of the improved version of the effective
potential, in the next section.

The effective potential was also calculated up to three loops, in
the minimal subtraction (MS) scheme, as follows,

\begin{align}
V_{\mathrm{eff}}^{U}\left(\sigma\right) & =\frac{\sigma^{2}}{2g\,\pi}S_{\mathrm{eff}}^{U}\left(\sigma;g,h,L\right),\label{eq:Eff. Potential}
\end{align}
with

\begin{align}
S_{\mathrm{eff}}^{U}\left(\sigma;g,h,L\right) & =A^{\left(0\right)}+A^{\left(1\right)}+A^{\left(3\right)}+\left(\frac{3}{16}\left(1-2N\right)Ng^{3}+\frac{3}{8}\left(1-2N\right)Ng^{2}h\right.\nonumber \\
 & \left.+\frac{3}{16}\left(1-2N\right)Ngh^{2}+Ng\right)L+\left(\frac{1}{8}N\left(2N+1\right)g^{3}+\frac{1}{4}\left(N-2N^{2}\right)g^{2}h\right.\nonumber \\
 & \left.+\frac{g^{2}}{2}N+\frac{1}{8}N\left(8N^{2}-6N+1\right)gh^{2}+\frac{1}{2}(1-2N)Ngh\right)L^{2}\nonumber \\
 & +\left(-\frac{1}{6}\left(N-2\right)Ng^{3}+\frac{2}{3}\left(1-2N\right)Ng^{2}h+\frac{1}{6}N\left(6N^{2}-7N+2\right)gh^{2}\right)L^{3},\label{eq: seff}
\end{align}
where 
\begin{equation}
L\equiv\ln\left[\frac{\sigma}{\mu}\right]\thinspace,\label{eq:defL}
\end{equation}
$\mu$ being the mass scale that is introduced to keep the dimensions
of the relevant quantities unchanged, and 
\begin{align}
A^{\left(0\right)} & =1,\,\,\,\,A^{\left(1\right)}=-\frac{N}{2}g,\nonumber \\
A^{\left(3\right)} & =-\frac{1}{96}\left(28\text{\ensuremath{\zeta}\ensuremath{\left(3\right)}}-9\right)N\left(2N-1\right)\left(gh^{2}+g^{3}+2g^{2}h\right),\label{eq:Tree leve coeficient}
\end{align}
where $\zeta\left(3\right)\simeq1.202$ is known as Apéri constant.

\section{\label{sec:Improvement-of-effective}Improvement of effective potential
for the GN model}

In this section we compute the improvement of effective potential
of the model defined by the Lagrangian\,(\ref{eq:GN modified model}).
We start with
\begin{align}
V_{\mathrm{eff}}^{I}\left(\sigma\right) & =\frac{\sigma^{2}}{2g\,\pi}S_{\mathrm{eff}}^{I}\left(\sigma;g,h,L\right),\label{eq:Improved Eff. Potential}
\end{align}
where $S_{\mathrm{eff}}^{\mathrm{I}}$ is a function that remains
to be determined. On dimensional grounds, we can assume the following
\emph{Ansatz},
\begin{align}
S_{\mathrm{eff}}^{I} & =A\left(g,h\right)+B\left(g,h\right)L+C\left(g,h\right)L^{2}+D\left(g,h\right)L^{3},\label{eq:improved Seff}
\end{align}
where $L$ is given by Eq.\,(\ref{eq:defL}), and the coefficients
$A$, $B$, $C$ and $D$ are functions only of the (dimensionless)
coupling constants. The main idea behind the method is the observation
that the coefficients in\,(\ref{eq:improved Seff}) are not all independent,
since changes in $\mu$ must be compensated for by changes in the
other parameters, according to the renormalization group. This is
the same as saying that the effective potential has to satisfy a RGE.
Following the procedure in\,\citep{Quinto2016,AHMADY2003221,CHISHTIE2007},
and using the conventions given in\,\citep{Luperini1991} and quoted
in the last section, we can write the RGE for $S_{\mathrm{eff}}$
in the form
\begin{align}
\left[-\left(1+\gamma_{\sigma}\left(g,h\right)\right)\frac{\partial}{\partial L}+\beta_{g}\left(g,h\right)\frac{\partial}{\partial g}+\beta_{h}\left(g,h\right)\frac{\partial}{\partial h}+\gamma_{R}\left(g,h\right)\right]S_{\mathrm{eff}}^{I}\left(\sigma;g,h,L\right) & =0,\label{eq:RGE}
\end{align}
where the renormalization group functions are defined by equations\,(\ref{eq:Beta g}),
(\ref{eq:Beta h}), (\ref{eq:Gamma sigma}) and\,(\ref{eq:Gamma R}).

One should note, at this point, that these functions were computed\,\citep{Luperini1991}
in the MS scheme. In principle, they should be adapted to a different
scheme for our applications -- however, as discussed in\,\citep{Quinto2016},
this is not necessary when UV divergences appear at second or higher
loop level, as it is the present case. Therefore, this issue does
not have to be dealt with and, for our purposes, we can directly apply
the renormalization group equations obtained in\,\citep{Luperini1991}
for the RGE improvement.

If we use the \emph{Ansatz} in Eq.\,(\ref{eq:improved Seff}) together
with Eq.\,(\ref{eq:RGE}), it is possible to calculate recursively,
order by order in the coupling constants, the functions $A\left(g,h\right)$,
$B\left(g,h\right)$, $C\left(g,h\right)$ and $D\left(g,h\right)$.
In particular, $A\left(g,h\right)$ is fixed by the tree-level effective
potential, Eq.\,(\ref{eq:Tree leve coeficient}), in the form
\begin{align}
A\left(g,h\right) & =A^{\left(0\right)}+A^{\left(1\right)}+A^{\left(3\right)},\label{eq:A funtion}
\end{align}
where $A^{\left(i\right)}$ with $i=0,1,3$ are known functions, and
again the superscript represents the global power of coupling constants
in each term. Following the same pattern, we want to calculate the
remaining functions order by order in coupling constants, so we also
write,
\begin{align}
B\left(g,h\right) & =B^{\left(0\right)}+B^{\left(1\right)}+B^{\left(2\right)}+B^{\left(3\right)}+B^{\left(4\right)}+B^{\left(5\right)}+B^{\left(6\right)}+\cdots\thinspace,\label{eq:B function}\\
C\left(g,h\right) & =C^{\left(0\right)}+C^{\left(1\right)}+C^{\left(2\right)}+C^{\left(3\right)}+C^{\left(4\right)}+C^{\left(5\right)}+C^{\left(6\right)}+\cdots\thinspace,\label{eq:C funtion}\\
D\left(g,h\right) & =D^{\left(0\right)}+D^{\left(1\right)}+D^{\left(2\right)}+D^{\left(3\right)}+D^{\left(4\right)}+D^{\left(5\right)}+D^{\left(6\right)}+\cdots\thinspace.\label{eq:D funtion}
\end{align}
Terms of $\mathcal{O}\left(L^{0}\right)$ in the RGE correspond to
the function $B\left(g,h\right)$ in the \emph{Ansatz}\,(\ref{eq:improved Seff}).
These can be calculated from our knowledge of $A\left(g,h\right)$
and the renormalization group functions. To do that, we substitute\,(\ref{eq:improved Seff})
into\,(\ref{eq:RGE}) and separate the terms proportional to $L^{0}$,
obtaining the following expression,
\begin{align}
-\left(1+\gamma_{\sigma}\left(g,h\right)\right)B\left(g,h\right)+\left[\beta_{g}\left(g,h\right)\frac{\partial}{\partial g}+\beta_{h}\left(g,h\right)\frac{\partial}{\partial h}+\gamma_{R}\left(g,h\right)\right]A\left(g,h\right) & =0.\label{eq:RGE for OL zero}
\end{align}
Substituting\,(\ref{eq:A funtion}) and\,(\ref{eq:B function}),
together with the renormalization group functions, Eqs.\,(\ref{eq:Beta g}),
(\ref{eq:Beta h}), (\ref{eq:Gamma sigma}) and\,(\ref{eq:Gamma R}),
into\,(\ref{eq:RGE for OL zero}) leads us to the following expression,

\begin{align}
-\left(1+\gamma_{\sigma}^{\left(1\right)}+\gamma_{\sigma}^{\left(2\right)}+\gamma_{\sigma}^{\left(3\right)}\right)\left(B^{\left(0\right)}+B^{\left(1\right)}+B^{\left(2\right)}+B^{\left(3\right)}+B^{\left(4\right)}+...\right)+\nonumber \\
+\left[\left(\beta_{g}^{\left(2\right)}+\beta_{g}^{\left(3\right)}+\beta_{g}^{\left(4\right)}\right)\frac{\partial}{\partial g}+\left(\beta_{h}^{\left(2\right)}+\beta_{h}^{\left(3\right)}+\beta_{h}^{\left(4\right)}\right)\frac{\partial}{\partial h}+\right.\nonumber \\
+\left.\gamma_{R}^{\left(1\right)}+\gamma_{R}^{\left(3\right)}\right]\left(A^{\left(0\right)}+A^{\left(1\right)}+A^{\left(3\right)}\right) & =0,
\end{align}
and from the previous equation, we can obtain,\begin{subequations}

\begin{align}
B^{\left(0\right)} & =0,\\
B^{\left(1\right)} & =-\gamma_{\sigma}^{\left(1\right)}B^{\left(0\right)}+\gamma_{R}^{\left(1\right)}A^{\left(0\right)}=Ng,\\
B^{\left(2\right)} & =-\gamma_{\sigma}^{\left(2\right)}B^{\left(0\right)}-\gamma_{\sigma}^{\left(1\right)}B^{\left(1\right)}+\left[\beta_{g}^{\left(2\right)}\frac{\partial}{\partial g}+\beta_{h}^{\left(2\right)}\frac{\partial}{\partial h}+\gamma_{R}^{\left(1\right)}\right]A^{\left(1\right)}=0,\\
B^{\left(3\right)} & =-\gamma_{\sigma}^{\left(3\right)}B^{\left(0\right)}-\gamma_{\sigma}^{\left(2\right)}B^{\left(1\right)}-\gamma_{\sigma}^{\left(1\right)}B^{\left(2\right)}+\left[\beta_{g}^{\left(3\right)}\frac{\partial}{\partial g}+\beta_{h}^{\left(3\right)}\frac{\partial}{\partial h}\right]A^{\left(1\right)}+\gamma_{R}^{\left(3\right)}A^{\left(0\right)}\nonumber \\
 & =\frac{3}{16}\left(1-2N\right)N\left\{ \,g^{3}+2\,g^{2}h+\,gh^{2}\right\} ,\\
B^{\left(4\right)} & =-\gamma_{\sigma}^{\left(3\right)}B^{\left(1\right)}-\gamma_{\sigma}^{\left(2\right)}B^{\left(2\right)}-\gamma_{\sigma}^{\left(1\right)}B^{\left(3\right)}+\left[\beta_{g}^{\left(4\right)}\frac{\partial}{\partial g}+\beta_{h}^{\left(4\right)}\frac{\partial}{\partial h}+\gamma_{R}^{\left(3\right)}\right]A^{\left(1\right)}+\left[\beta_{g}^{\left(2\right)}\frac{\partial}{\partial g}+\beta_{h}^{\left(2\right)}\frac{\partial}{\partial h}+\gamma_{R}^{\left(1\right)}\right]A^{\left(3\right)}\nonumber \\
 & =\frac{1}{48}N\left(2N-1\right)\left\{ \left[-42\text{\ensuremath{\zeta}\ensuremath{\left(3\right)}}+\left(28\text{\ensuremath{\zeta}\ensuremath{\left(3\right)}}-9\right)N+9\right]g^{4}+\left[-126\text{\ensuremath{\zeta}\ensuremath{\left(3\right)}}+\left(112\text{\text{\ensuremath{\zeta}\ensuremath{\left(3\right)}}}-27\right)N+27\right]g^{3}h\right.\nonumber \\
 & \left.+\left[-126\text{\text{\ensuremath{\zeta}\ensuremath{\left(3\right)}}}+\left(140\text{\text{\ensuremath{\zeta}\ensuremath{\left(3\right)}}}-27\right)N+27\right]g^{2}h^{2}+\left[-42\text{\ensuremath{\zeta}\ensuremath{\left(3\right)}}+\left(56\text{\ensuremath{\zeta}\ensuremath{\left(3\right)}}-9\right)N+9\right]gh^{3}\right\} ,
\end{align}
\begin{align}
B^{\left(5\right)} & =-\gamma_{\sigma}^{\left(3\right)}B^{\left(2\right)}-\gamma_{\sigma}^{\left(2\right)}B^{\left(3\right)}-\gamma_{\sigma}^{\left(1\right)}B^{\left(4\right)}+\left[\beta_{g}^{\left(3\right)}\frac{\partial}{\partial g}+\beta_{h}^{\left(3\right)}\frac{\partial}{\partial h}\right]A^{\left(3\right)}\nonumber \\
 & =-\frac{7}{96}\text{\ensuremath{\zeta}\ensuremath{\left(3\right)}}\left(4N^{2}-1\right)g^{5}-N\left(2N-1\right)\left\{ \frac{1}{48}\left(14\text{\text{\ensuremath{\zeta}\ensuremath{\left(3\right)}}}+\left(28\text{\ensuremath{\zeta}\ensuremath{\left(3\right)}}-9\right)N^{2}+\left(9-28\text{\text{\ensuremath{\zeta}\ensuremath{\left(3\right)}}}\right)N\right)g^{4}h\right.\nonumber \\
 & +\frac{1}{48}\left(21\text{\text{\ensuremath{\zeta}\ensuremath{\left(3\right)}}}+\left(112\text{\ensuremath{\zeta}\ensuremath{\left(3\right)}}-27\right)N^{2}-9\left(14\text{\text{\ensuremath{\zeta}\ensuremath{\left(3\right)}}}-3\right)N\right)g^{3}h^{2}\nonumber \\
 & +\frac{1}{48}\left(14\text{\ensuremath{\zeta}\ensuremath{\left(3\right)}}+\left(140\text{\ensuremath{\zeta}\ensuremath{\left(3\right)}}-27\right)N^{2}+\left(27-140\text{\text{\ensuremath{\zeta}\ensuremath{\left(3\right)}}}\right)N\right)g^{2}h^{3}\nonumber \\
 & \left.+\frac{1}{96}\left(7\text{\ensuremath{\zeta}\ensuremath{\left(3\right)}}+2\left(56\text{\text{\ensuremath{\zeta}\ensuremath{\left(3\right)}}}-9\right)N^{2}+\left(18-98\text{\text{\ensuremath{\zeta}\ensuremath{\left(3\right)}}}\right)N\right)gh^{4}\right\} ,
\end{align}

\begin{align}
B^{\left(6\right)} & =-\gamma_{\sigma}^{\left(3\right)}B^{\left(3\right)}-\gamma_{\sigma}^{\left(2\right)}B^{\left(4\right)}-\gamma_{\sigma}^{\left(1\right)}B^{\left(5\right)}+\left[\beta_{g}^{\left(4\right)}\frac{\partial}{\partial g}+\beta_{h}^{\left(4\right)}\frac{\partial}{\partial h}+\gamma_{R}^{\left(3\right)}\right]A^{\left(3\right)},\nonumber \\
 & =N\left(2N-1\right)\left\{ \frac{1}{768}\left(-182\text{\ensuremath{\zeta}\ensuremath{\left(3\right)}}+2\left(56\text{\ensuremath{\zeta}\ensuremath{\left(3\right)}}-9\right)N^{2}+\left(28\text{\ensuremath{\zeta}\ensuremath{\left(3\right)}}-27\right)N+45\right)g^{6}\right.\nonumber \\
 & +\frac{1}{768}\left(-910\text{\ensuremath{\zeta}\ensuremath{\left(3\right)}}+2(196\text{\ensuremath{\zeta}\ensuremath{\left(3\right)}}-9)N^{2}+3(224\text{\ensuremath{\zeta}\ensuremath{\left(3\right)}}-69)N+225\right)g^{5}h\nonumber \\
 & +\frac{1}{384}\left(-910\ensuremath{\zeta}\left(3\right)+8\left(28\ensuremath{\zeta}\left(3\right)-9\right)N^{3}-42\left(8\ensuremath{\zeta}\left(3\right)-3\right)N^{2}+\left(1204\ensuremath{\zeta}\left(3\right)-279\right)N+225\right)g^{4}h^{2}\nonumber \\
 & +\frac{1}{384}\left(-910\ensuremath{\zeta}\left(3\right)+8\left(112\ensuremath{\zeta}\left(3\right)-27\right)N^{3}+\left(342-1624\ensuremath{\zeta}\left(3\right)\right)N^{2}+\left(1736\ensuremath{\zeta}\left(3\right)-351\right)N+225\right)g^{3}h^{3}\nonumber \\
 & +\frac{1}{768}\left(-910\ensuremath{\zeta}\left(3\right)+16\left(140\ensuremath{\zeta}\left(3\right)-27\right)N^{3}+\left(630-3472\text{\ensuremath{\zeta}\ensuremath{\left(3\right)}}\right)N^{2}+9\left(252\ensuremath{\zeta}\left(3\right)-47\right)N+225\right)g^{2}h^{4}\nonumber \\
 & \left.+\frac{1}{768}\left(-182\ensuremath{\zeta}\left(3\right)+16\left(56\ensuremath{\zeta}\left(3\right)-9\right)N^{3}-6\left(196\ensuremath{\zeta}\left(3\right)-33\right)N^{2}+\left(560\ensuremath{\zeta}\left(3\right)-99\right)N+45\right)gh^{5}\right\} .
\end{align}
\end{subequations}For the purpose of this paper, we will only consider
terms up to sixth order in the coupling constants because we only
know the $\beta$ function up to four order.

Terms of $\mathcal{O}\left(L\right)$ in the RGE will lead to the
calculation of the $C$'s in\,(\ref{eq:C funtion}) from the knowledge
we already have from the perturbative calculations, as well as the
$B$'s that we just obtained. Repeating the same procedure as before,
we find the following results,\begin{subequations}

\begin{align}
C^{\left(0\right)} & =C^{\left(1\right)}=0,\\
C^{\left(2\right)} & =\frac{N}{2}g^{2}+\frac{1}{2}\left(N-2N^{2}\right)gh\\
C^{\left(3\right)} & =\frac{1}{8}N\left(2N+1\right)g^{3}+\frac{1}{4}\left(N-2N^{2}\right)g^{2}h+\frac{1}{8}N\left(8N^{2}-6N+1\right)gh^{2},\\
C^{\left(4\right)} & =\frac{1}{8}N\left(2N^{2}-5N+3\right)g^{4}+\frac{1}{8}N\left(8N^{2}-22N+9\right)g^{3}h+\frac{1}{8}N\left(22N^{2}-29N+9\right)g^{2}h^{2}\nonumber \\
 & +\frac{1}{8}N\left(-8N^{3}+16N^{2}-12N+3\right)gh^{3},
\end{align}
\begin{align}
C^{\left(5\right)} & =\frac{1}{384}N\left(21\left(32\text{\ensuremath{\zeta}\ensuremath{\left(3\right)}}-5\right)-24\left(28\ensuremath{\zeta}\left(3\right)-9\right)N^{3}+224\left(10\text{\ensuremath{\zeta}\ensuremath{\left(3\right)}}-3\right)N^{2}+\left(540-2296\text{\ensuremath{\zeta}\ensuremath{\left(3\right)}}\right)N\right)g^{5}\nonumber \\
 & -N\left(2N-1\right)\left\{ \frac{1}{96}\left(21\left(32\ensuremath{\zeta}\left(3\right)-5\right)+8\left(56\ensuremath{\zeta}\left(3\right)-15\right)N^{2}-4\left(287\ensuremath{\zeta}\left(3\right)-66\right)N\right)g^{4}h\right.\nonumber \\
 & +\frac{1}{64}\left(21\left(32\ensuremath{\zeta}\left(3\right)-5\right)+8\left(77\ensuremath{\zeta}\left(3\right)-15\right)N^{2}-6\left(224\ensuremath{\zeta}\left(3\right)-33\right)N\right)g^{3}h^{2}\nonumber \\
 & +\frac{1}{96}\left(21\left(32\text{\ensuremath{\zeta}\ensuremath{\left(3\right)}}-5\right)+12\left(70\text{\ensuremath{\zeta}\ensuremath{\left(3\right)}}-1\right)N^{2}-44\left(35\ensuremath{\zeta}\left(3\right)-3\right)N\right)g^{2}h^{3}\nonumber \\
 & \left.-\frac{1}{384}\left(-672\ensuremath{\zeta}\left(3\right)+192N^{3}-4\left(280\ensuremath{\zeta}\left(3\right)+51\right)N^{2}+2\left(868\ensuremath{\zeta}\left(3\right)-33\right)N+105\right)gh^{4}\right\} ,
\end{align}
\begin{align}
C^{\left(6\right)} & =\frac{1}{1536}N\left(448\ensuremath{\zeta}\left(3\right)+96\left(14\ensuremath{\zeta}\left(3\right)-1\right)N^{3}+\left(60-2016\ensuremath{\zeta}\left(3\right)\right)N^{2}-32\left(7\ensuremath{\zeta}\left(3\right)+6\right)N+141\right)g^{6}\nonumber \\
 & +N\left(2N-1\right)\left\{ \frac{1}{1536}\left(-5\left(448\ensuremath{\zeta}\left(3\right)+141\right)+96\left(28\ensuremath{\zeta}\left(3\right)-9\right)N^{3}+\left(2244-5152\ensuremath{\zeta}\left(3\right)\right)N^{2}\right.\right.\nonumber \\
 & \left.+\left(4816\ensuremath{\zeta}\left(3\right)-996\right)N\right)g^{5}h+\frac{1}{768}\left(-5\left(448\text{\ensuremath{\zeta}\ensuremath{\left(3\right)}}+141\right)+24\left(308\text{\ensuremath{\zeta}\ensuremath{\left(3\right)}}-87\right)N^{3}\right.\nonumber \\
 & \left.+\left(4656-18032\text{\ensuremath{\zeta}\ensuremath{\left(3\right)}}\right)N^{2}+2\left(6496\text{\ensuremath{\zeta}\ensuremath{\left(3\right)}}-771\right)N\right)g^{4}h^{2}\nonumber \\
 & +\frac{1}{768}\left(-5\left(448\text{\ensuremath{\zeta}\ensuremath{\left(3\right)}}+141\right)+40\left(392\text{\ensuremath{\zeta}\ensuremath{\left(3\right)}}-87\right)N^{3}+\left(6036-35280\text{\ensuremath{\zeta}\ensuremath{\left(3\right)}}\right)N^{2}\right.\nonumber \\
 & \left.+72\left(294\ensuremath{\zeta}\left(3\right)-29\right)N\right)g^{3}h^{3}+\frac{1}{1536}\left(-5\left(448\ensuremath{\zeta}\left(3\right)+141\right)+16\left(1820\ensuremath{\zeta}\left(3\right)-183\right)N^{3}\right.\nonumber \\
 & \left.-112\left(508\ensuremath{\zeta}\left(3\right)-57\right)N^{2}+\left(29344\ensuremath{\zeta}\left(3\right)-2634\right)N\right)g^{2}h^{4}-\frac{1}{1536}\left(448\text{\ensuremath{\zeta}\ensuremath{\left(3\right)}}+768N^{4}\right.\nonumber \\
 & \left.\left.-112\left(88\text{\ensuremath{\zeta}\ensuremath{\left(3\right)}}+3\right)N^{3}+4\left(4144\text{\ensuremath{\zeta}\ensuremath{\left(3\right)}}-285\right)N^{2}+\left(636-7504\text{\ensuremath{\zeta}\ensuremath{\left(3\right)}}\right)N+141\right)gh^{5}\right\} ,
\end{align}
\end{subequations}

Going to $\mathcal{O}\left(L^{2}\right)$ in the RGE, we can find
all the $D$\textasciiacute s in\,(\ref{eq:D funtion}), and the
result is,\begin{subequations}

\begin{align}
D^{\left(0\right)} & =D^{\left(1\right)}=D^{\left(2\right)}=0\\
D^{\left(3\right)} & =-\frac{1}{6}\left(N-2\right)Ng^{3}+\frac{2}{3}\left(1-2N\right)Ng^{2}h+\frac{1}{6}N\left(6N^{2}-7n+2\right)gh^{2}\\
D^{\left(4\right)} & =\frac{1}{6}\left(-N^{3}+2N^{2}+N\right)g^{4}-\frac{1}{4}N\left(2N^{2}+3N-2\right)g^{3}h+\frac{1}{2}N\left(6N^{2}-5N+1\right)g^{2}h^{2}\nonumber \\
 & +\frac{1}{12}N\left(-28N^{3}+40N^{2}-17N+2\right)gh^{3}\\
D^{\left(5\right)} & =\frac{1}{96}N\left(-24N^{3}+94N^{2}-107N+57\right)g^{5}+\frac{1}{48}N\left(-48N^{3}+206N^{2}-319N+114\right)g^{4}h\nonumber \\
 & +\frac{1}{16}N\left(-30N^{3}+211N^{2}-212N+57\right)g^{3}h^{2}+\frac{1}{48}N\left(-468N^{3}+836N^{2}-529N+114\right)g^{2}h^{3}\nonumber \\
 & +\frac{1}{96}N\left(384N^{4}-828N^{3}+724N^{2}-317N+57\right)gh^{4},
\end{align}
\begin{align}
D^{\left(6\right)} & =\frac{1}{288}N\left(840\text{\ensuremath{\zeta\left(3\right)}}+24\left(28\zeta\left(3\right)-9\right)N^{4}+\left(948-3080\zeta\left(3\right)\right)N^{3}+52\left(98\zeta\left(3\right)-27\right)N^{2}\right.\nonumber \\
 & \left.+\left(786-3542\zeta\left(3\right)\right)N-93g^{6}\right)+N\left(2N-1\right)\left\{ \frac{1}{96}\left(-1400\zeta\left(3\right)+2\left(364\zeta\left(3\right)-99\right)N^{3}\right.\right.\nonumber \\
 & \left.+\left(724-2828\zeta\left(3\right)\right)N^{2}+\left(3542\zeta\left(3\right)-766\right)N+155\right)g^{5}h+\frac{1}{144}\left(-4200\zeta\left(3\right)\right.\nonumber \\
 & \left.+15\left(196\zeta\left(3\right)-45\right)N^{3}+\left(2118-10612\zeta\left(3\right)\right)N^{2}+14\left(853\zeta\left(3\right)-114\right)N+465\right)g^{4}h^{2}\nonumber \\
 & +\frac{1}{144}\left(-4200\zeta\left(3\right)+\left(4116\zeta\left(3\right)-699\right)N^{3}-22\left(602\zeta\left(3\right)-39\right)N^{2}+2\left(6629\zeta\left(3\right)-447\right)N\right.\nonumber \\
 & \left.+465\right)g^{3}h^{3}+\frac{1}{96}\left(-1400\text{\ensuremath{\zeta\left(3\right)}}+2\left(980\zeta\left(3\right)+263\right)N^{3}-4\left(1365\zeta\left(3\right)+134\right)N^{2}\right.\nonumber \\
 & \left.+\left(4858\text{\ensuremath{\zeta\left(3\right)}}-64\right)N+155\right)g^{2}h^{4}-\frac{1}{288}\left(840\zeta\left(3\right)+864N^{4}-1680\left(\text{\ensuremath{\zeta\left(3\right)}}+1\right)N^{3}\right.\nonumber \\
 & \left.\left.+44\left(91\zeta\left(3\right)+24\right)N^{2}-2\left(1589\zeta\left(3\right)+51\right)N-93\right)gh^{5}\right\} .
\end{align}
\end{subequations}

Finally, with the values of $A,B,C$ and $D$ that have been obtained,
we obtain $V_{\mathrm{eff}}^{I}\left(\sigma\right)$, which we call
the improved effective potential, since it contains higher-orders
(in the coupling constants) terms that were obtained from the RGE,
and beyond what can be obtained by direct loop calculation, as presented
in Sec.\,\ref{sec:2}. Notice that it is possible to obtain the unimproved
version of the effective potential, $V_{\mathrm{eff}}^{U}\left(\sigma\right)$,
that was calculated up to three loop order in\,\citep{Luperini1991}
by setting $B_{4}=B_{5}=B_{6}=0$, $C_{4}=C_{5}=C_{6}=0$ and $D_{4}=D_{5}=D_{6}=0$.
This is a proof of the consistency of our calculations.

\section{\label{sec:Dynamical-symmetric-breaking}Dynamical symmetric breaking}

We start this section analyzing the behavior of the DSB for the unimproved
and improved versions of the effective potential, Eq.\,(\ref{eq:Eff. Potential}).
First, one has to recognize that the effective potentials that we
computed actually correspond to the \emph{regularized} effective potential,
and we still need to fix a finite renormalization constant that is
introduced via
\begin{align}
V_{\mathrm{eff},\boldsymbol{R}}^{U/I}\left(\sigma\right)= & V_{\mathrm{eff}}^{U/I}\left(\sigma\right)+\sigma^{2}\rho,\label{eq:Regularized Im-effective potential}
\end{align}
where $\rho$ can be fixed with the Coleman-Weinberg (CW)\,\citep{Coleman1973}
condition, 
\begin{align}
\left.\frac{d^{2}}{d\sigma^{2}}V_{\mathrm{eff,}\boldsymbol{R}}^{U/I}\left(\sigma\right)\right|_{\sigma=\mu} & =\frac{1}{g}.\label{eq:CW condition}
\end{align}

The second step is to enforce that $V_{\mathrm{eff,\boldsymbol{R}}}^{U/I}\left(\sigma\right)$
has a minimum at $\sigma=\mu$. This is done imposing the condition,
\begin{align}
\left.\frac{d}{d\sigma}V_{\mathrm{eff}\boldsymbol{R}}^{U/I}\left(\sigma\right)\right|_{\sigma=\mu} & =0,\label{eq:Minima condition of Improved potential}
\end{align}
together with
\begin{align}
m_{\sigma}^{2} & =\left.\frac{d^{2}}{d\sigma^{2}}V_{\mathrm{eff,}\boldsymbol{R}}^{U/I}\left(\sigma\right)\right|_{\sigma=\mu}=\frac{1}{g}>0,\label{eq:Mass of sigma}
\end{align}
where $m_{\sigma}^{2}$ is the mass generated by radiative corrections
for the $\sigma$ scalar field. It is interesting to notice that,
here, this last condition\,(\ref{eq:Mass of sigma}) is actually
equivalent to the CW condition, that is to say, Eq.\,\,(\ref{eq:Mass of sigma})
is automatically satisfied once\,(\ref{eq:CW condition}) is enforced.
The same does not happen in other models that were studied within
this approach, such as\,\citep{Elias2003,Dias2010,Quinto2016}, where
Eq.\,\,(\ref{eq:Mass of sigma}) provides an additional selection
rule to be considered when looking into solutions for Eq.\,(\ref{eq:Minima condition of Improved potential}).

From a computational point of view, since we want to study the general
properties of the DSB mechanism in this model for a wide range of
the values of its coupling constants, we will use Eq.\,(\ref{eq:Minima condition of Improved potential})
to fix the value of the constant $g^{I}$ as a function of $h$ and
$N$, which will remain as free parameters. Also, at this point, the
rescaling $g\rightarrow g/\pi$ and $h\rightarrow h/\pi$ suggested
in\,\citep{Luperini1991} was implemented. Upon explicit calculation,
Eq.\,(\ref{eq:Minima condition of Improved potential}) turns out
to be a polynomial equation in $g^{I}$, and among its solutions we
look for those which are real and positive, and which lie in the perturbative
regime, $g<1$.

To analyze the DSB in our model both for the unimproved and improved
cases, we created a program in \emph{Mathematica©} to systematically
apply the previous steps for arbitrary values of the free parameters.
In other words: for any reasonable value of $h$ and $N$, we apply
the CW condition, Eq.\,(\ref{eq:CW condition}), to fix the renormalization
constant $\rho$, then we use Eq.\,(\ref{eq:Minima condition of Improved potential})
to find solutions of $g$ in terms of $h$ and $N$, from which we
separate the physical solutions that are real and positive, and also
satisfy $g<1$ to ensure we are within the perturbative regime. Any
solution with $g>1$ is discarded as nonphysical, since our approach
is inherently perturbative. This procedure is applied both for the
unimproved and improved regularized effective potentials, for the
sake of comparing both cases, and we denote as $g^{I}$ the value
of the coupling constant $g$ obtained with the improved potential,
and $g^{U}$ with the unimproved one.

\begin{figure}[H]
\begin{centering}
\subfloat[\label{fig:Region plot A}]{\centering{}\includegraphics[scale=0.4]{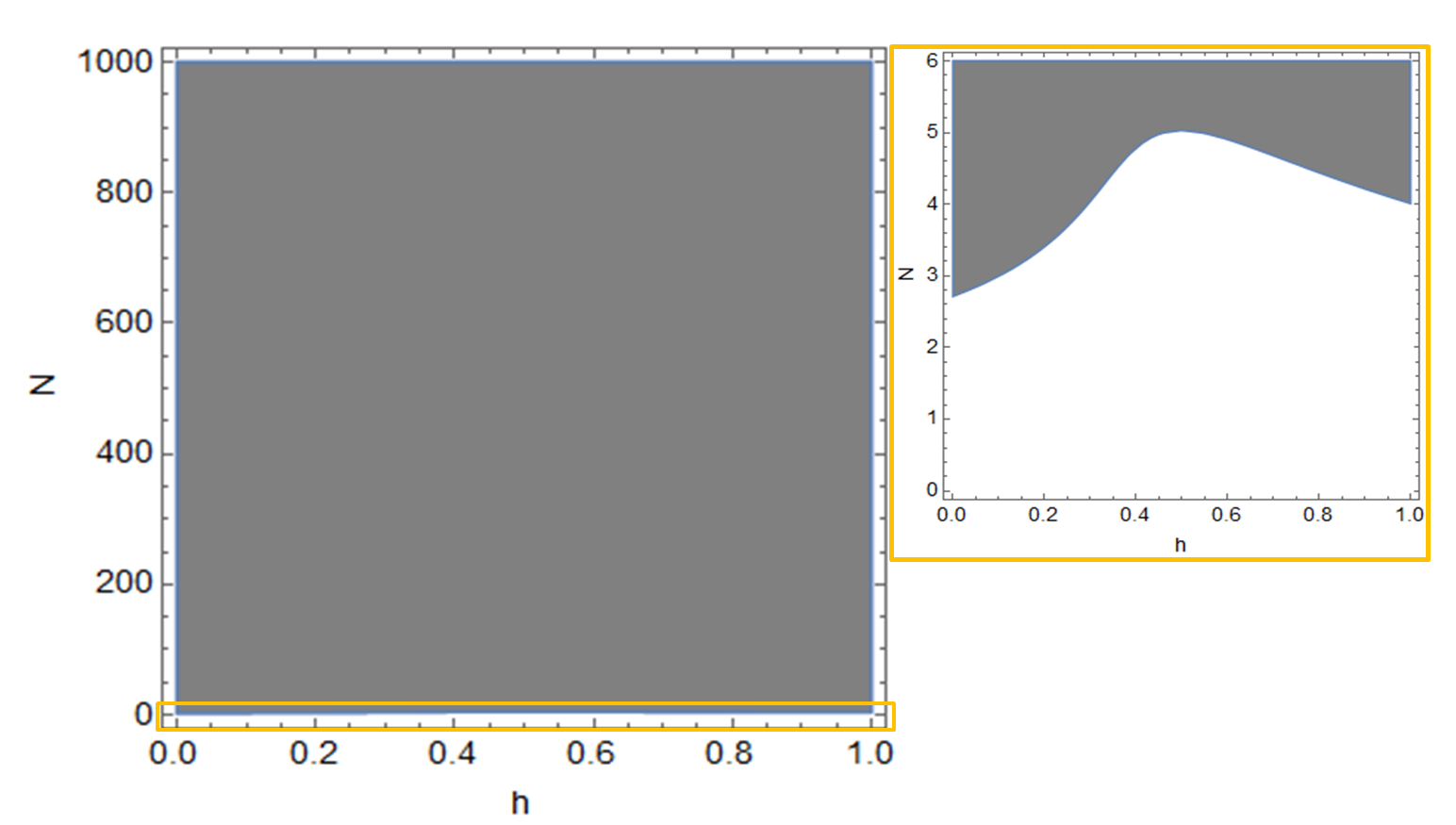}}
\par\end{centering}
\centering{}\subfloat[\label{fig:Region plot B}]{\centering{}\includegraphics[scale=0.4]{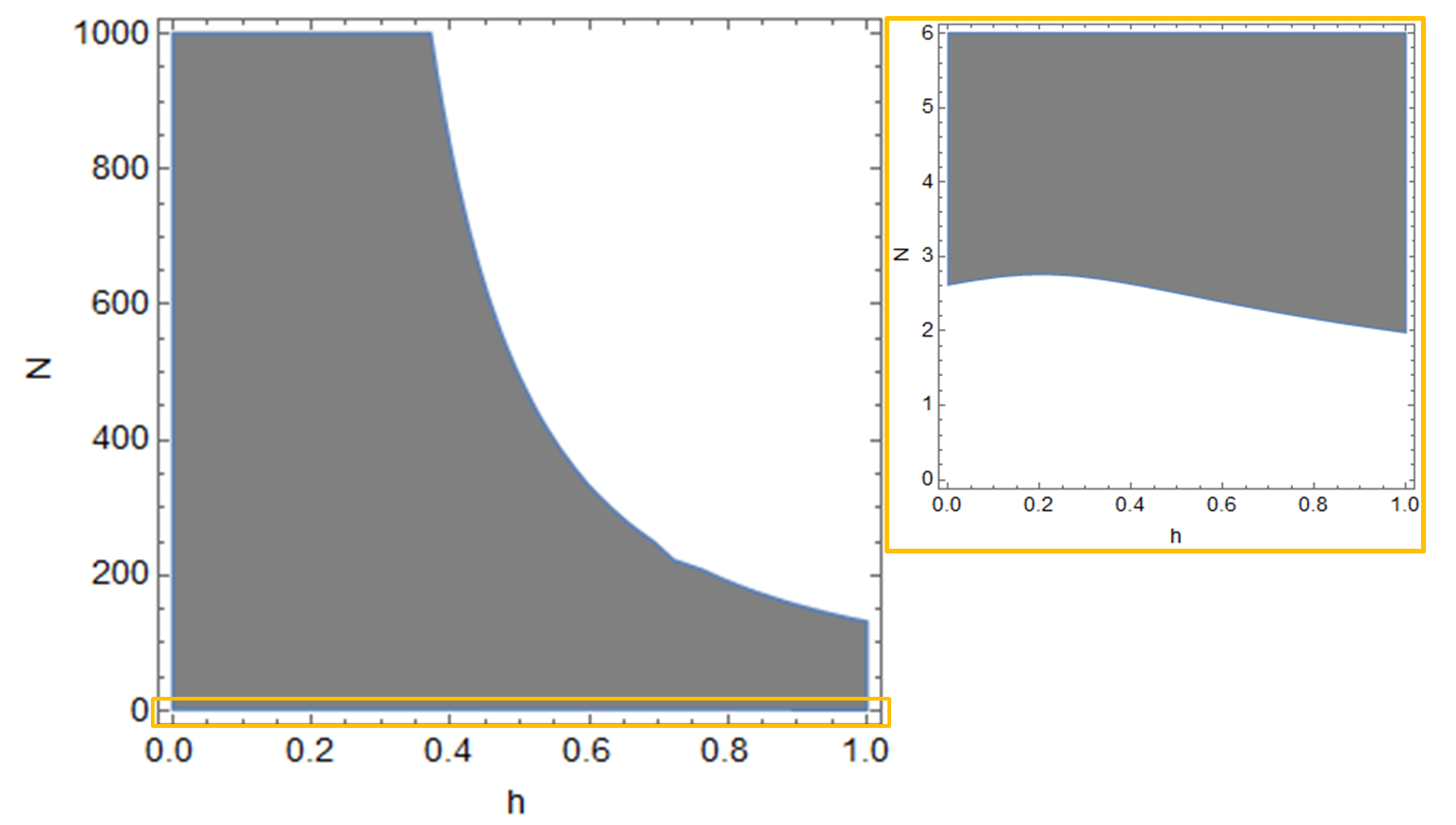}}\caption{\label{fig:Region plot}The region plots of $N$ vs. $h$ showing
the parameter space in which it is possible to find the values of
the improved and unimproved coupling constants, $g^{I}$ and $g^{U}$,
respectively, in which the effective potential leads to consistent
perturbative DSB. The figure\,\ref{fig:Region plot A} is for the
unimproved case and the figure\,\ref{fig:Region plot B} is for the
improved case.}
\end{figure}

As a first step, the parameters space in which the DSB is operational
was found by scanning the whole parameter space determined by $0\leq h\leq1$
and $0\leq N\leq1000$, and obtaining a region plot showing where
the DSB occurs (i.e., the region where the previously explained procedure
yields consistent minima away from the origin). These plots are presented
in the figure\,\ref{fig:Region plot}, both for the unimproved (figure\,\ref{fig:Region plot A})
and improved (figure\,\ref{fig:Region plot B}) cases. As we can
see, the parameter space for which the DSB is possible in the improved
case is much smaller than the unimproved one, which is consistent
with previous results in this type of studies, for example in three
and four dimensional space-time models\,\citep{Quinto2016,Dias2014}.

\begin{figure}[H]
\begin{centering}
\subfloat[\label{fig:Coupling relation A}]{\centering{}\includegraphics[scale=0.25]{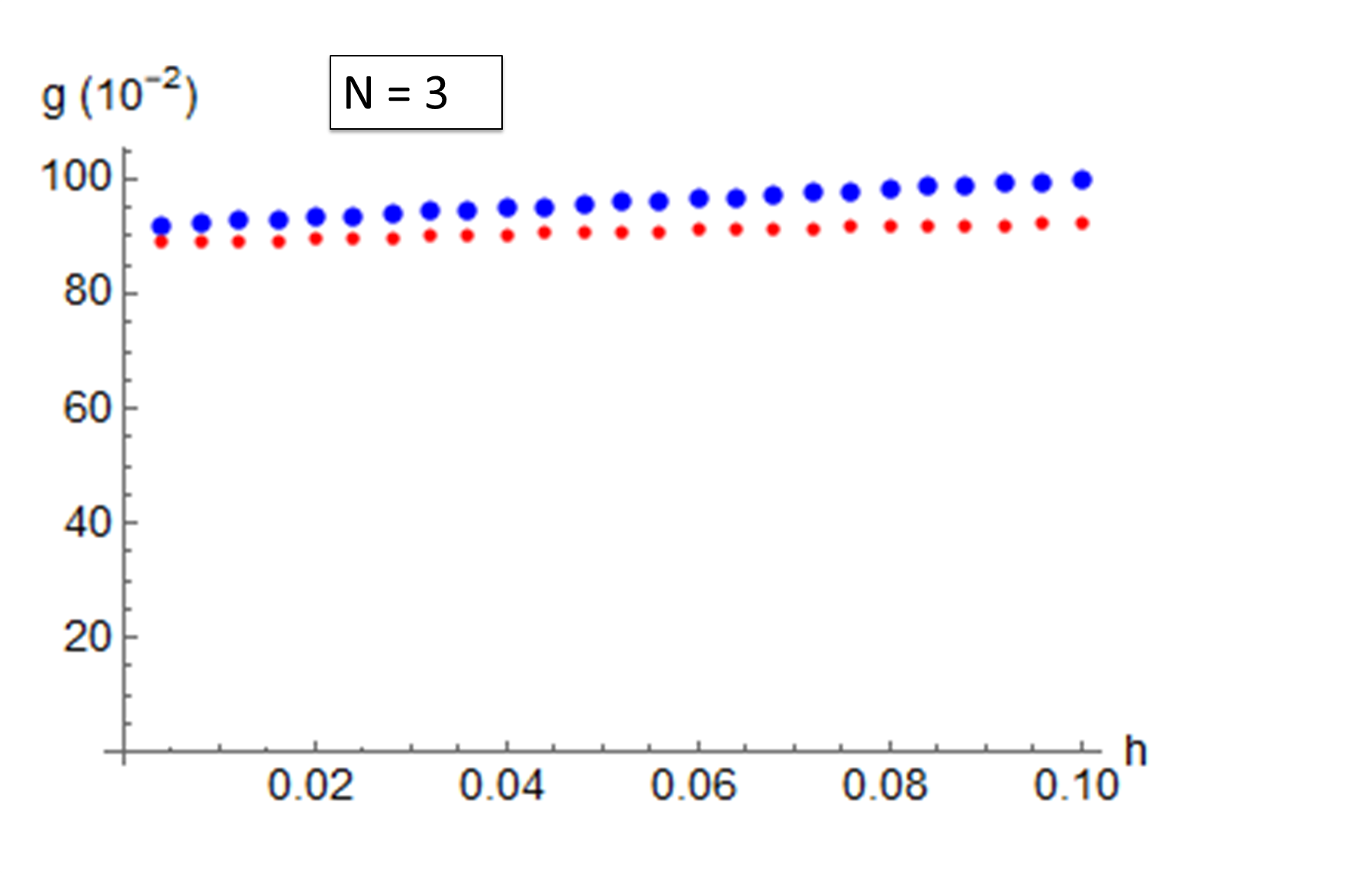}}~~\subfloat[\label{fig:Coupling relation B}]{\centering{}\includegraphics[scale=0.25]{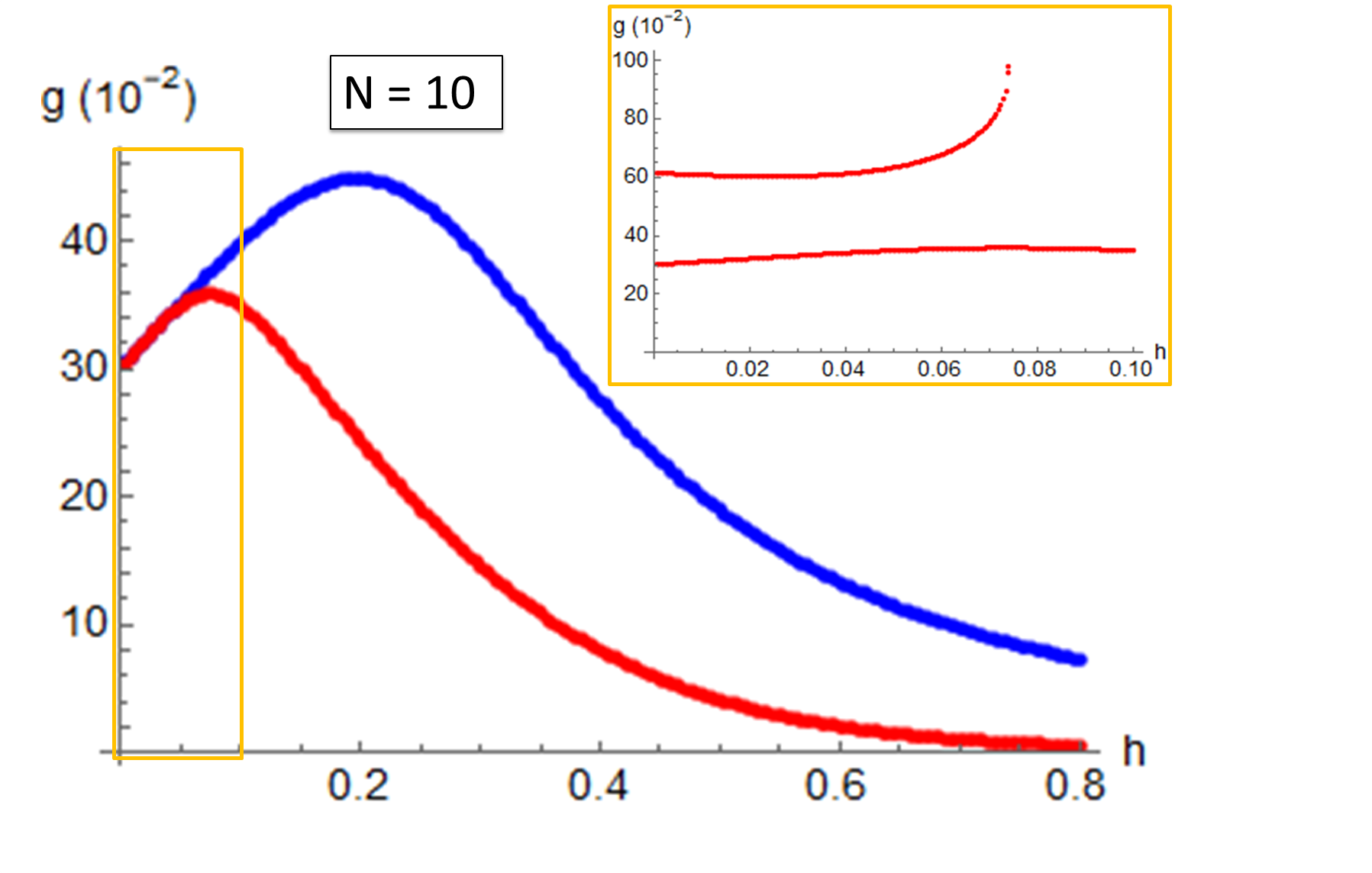}}
\par\end{centering}
\centering{}\subfloat[\label{fig:Coupling relation C}]{\centering{}\includegraphics[scale=0.25]{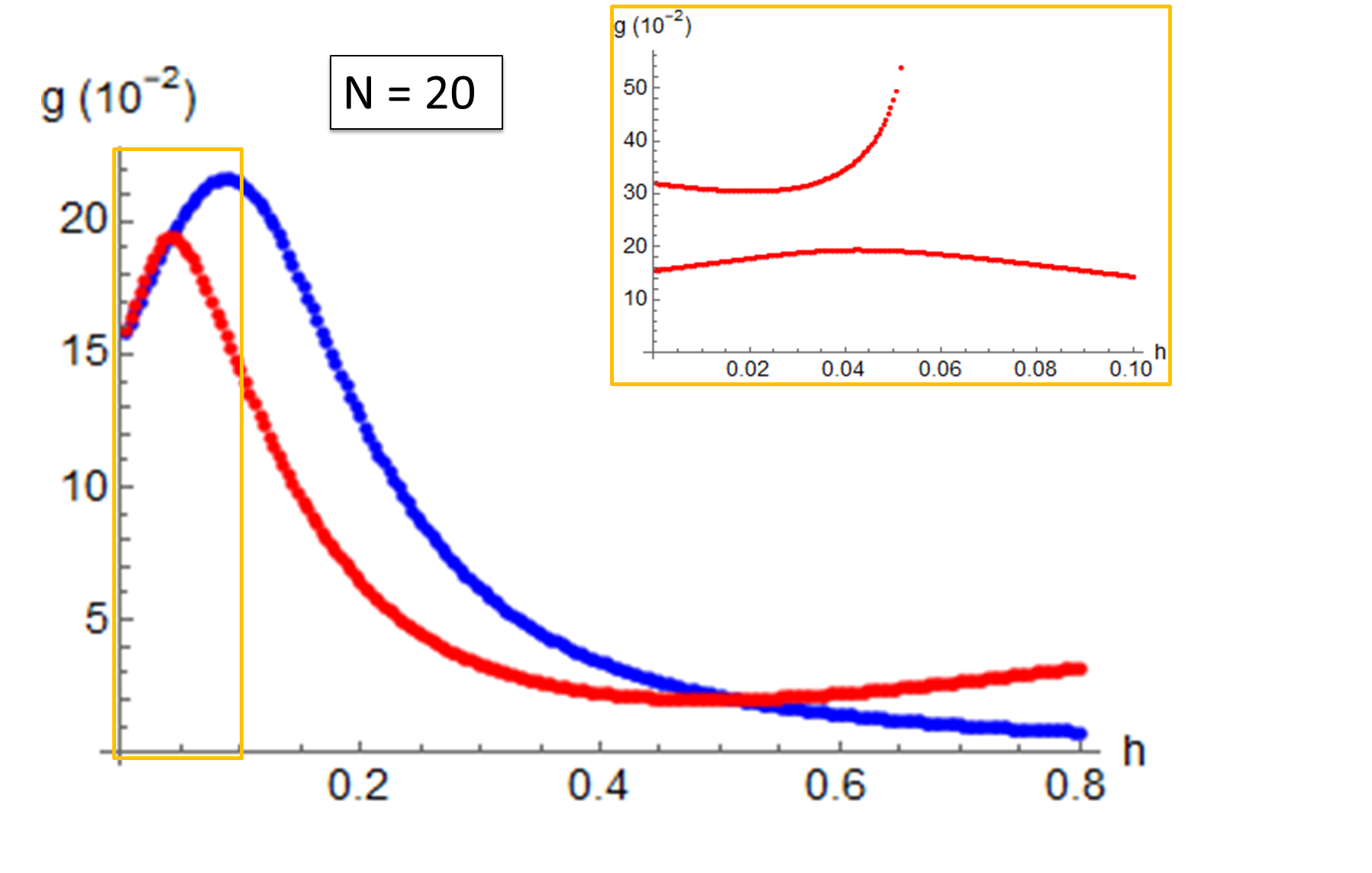}}~~\subfloat[\label{fig:Coupling relation D}]{\centering{}\includegraphics[scale=0.25]{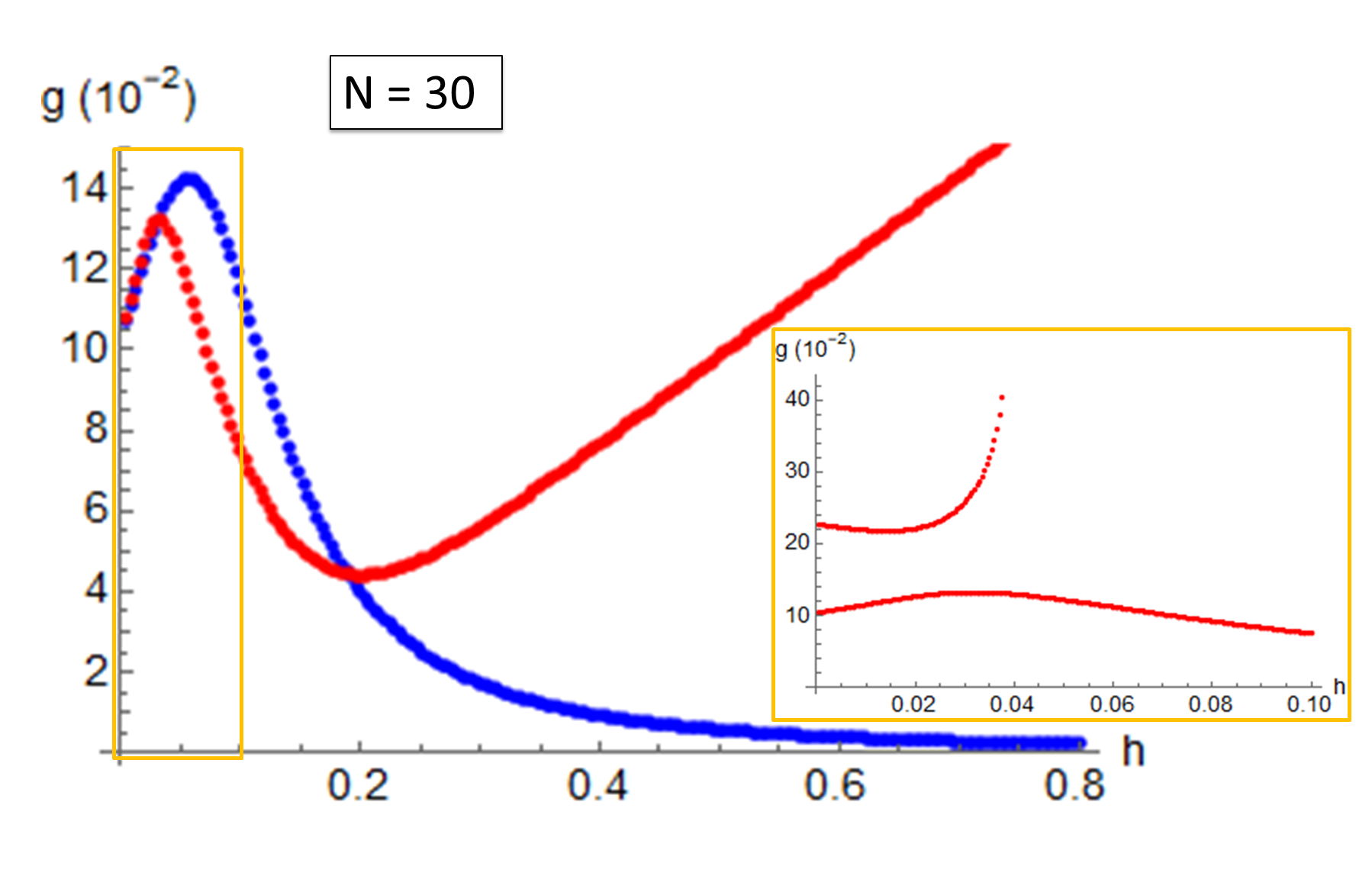}}\caption{\label{fig:Graph of coupling constants}The graphs of $g\left(10^{-2}\right)$
vs. $h$ represent the behavior of the improved, $g^{I}$ (red line),
and unimproved, $g^{U}$ (blue line) coupling constants as a function
of $h$ for certain values of $N$ ($N=3$, $N=10$, $N=20$ and $N=30$).
In figure\,\ref{fig:Coupling relation A}, the graph that describes
the behavior of the coupling constants as a function of $h$ is shown
for the case $N=3$, we can note that in this particular case, there
are only unique solutions for the values of $g^{I}$ and $g^{U}$
for an interval of values of $0\protect\leq h\protect\leq0.1$. On
the other hand, in the graphs presented in the figures\,\ref{fig:Coupling relation B}
- \ref{fig:Coupling relation D}, we observe two possible values for
$g^{I}$, where the second solution only appears for small values
of $h$ and decrease as $N$ increases.}
\end{figure}

We also observed the existence of more than one possible solution
for $g^{I}$ and $g^{U}$ for a given value of $h$ and $N$. For
this, the coupling constants for both cases were plotted as a function
of $h$ in the interval of $0\leq h\leq0.8$ for different fixed values
of $N$, as shown in figure\,\ref{fig:Graph of coupling constants}.

\begin{figure}[H]
\centering{}\includegraphics[scale=0.5]{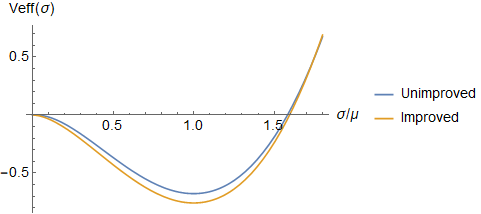}\caption{\label{fig:Plot of effective potential}The graph of $V_{\mathrm{eff}}^{U/I}\left(\sigma\right)$
vs. $\sigma/\mu$, where the behaviors of the minimum of improved
(yellow) and unimproved (blue) effective potentials are compared for
the values of $g^{I}=0.9073$ and $g^{U}=0.9579$, respectively, which
were obtained from the region plot, figure\,\ref{fig:Region plot},
when we take the values of $h=0.05$ and $N=3$. These values are
replaced in the unimproved and improved effective potentials, Eqs.\,(\ref{eq:Eff. Potential})
and\,(\ref{eq:Improved Eff. Potential}) respectively, and these
are evaluated in the interval $0\protect\leq\sigma/\mu\protect\leq1.8$.}
\end{figure}

It is interesting to note that for $N=3$ (figure\,\ref{fig:Coupling relation A}),
there is a single solution for $g$ in both cases. Also, we note that
there is a very small difference between $g^{I}$ and $g^{U}$ for
an interval of values of $0\leq h\leq0.1$. However, we can consider
an example where it is possible to observe the behavior between the
minimum of improved and unimproved effective potentials, for the values
of $g^{I}=0.9073$ and $g^{U}=0.9579$ respectively, corresponding
to $N=3$ and $h=0.05$, as we shown in the figure\,\ref{fig:Plot of effective potential}.

\begin{figure}[H]
\centering{}\includegraphics[scale=0.5]{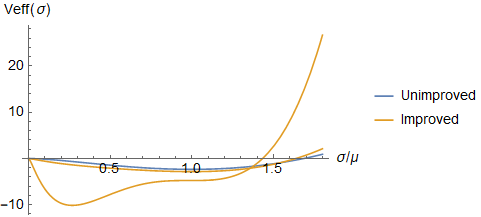}\caption{\label{fig:Plot of multipotentials}The graph of $V_{\mathrm{eff}}^{U/I}\left(\sigma\right)$
vs. $\sigma/\mu$, where the minima of the effective potentials are
compared for the unimproved (blue) and improved (yellow) cases. We
can see the presence of two improved and one unimproved effective
potentials, this is due to the presence of two possible solutions
for the improved case (see figure\,(\ref{fig:Coupling relation B})),
$g_{1}^{I}=0.6771$ and $g_{2}^{I}=0.3554$, and a single solution
for the unimproved case $g^{U}=0.3610$. In both cases, these solutions
are related to the minimum of the effective potential for values of
$N=10$ and $h=0.06$. These values were substituted in the Eqs.\,(\ref{eq:Eff. Potential})
and\,(\ref{eq:Improved Eff. Potential}) which were evaluated in
the interval $0\protect\leq\sigma/\mu\protect\leq1.8$.}
\end{figure}

On the other hand, if we analyze the cases $N=10$, $N=20$, and $N=30$,
which are shown in figures\,\ref{fig:Coupling relation B} to \ref{fig:Coupling relation D},
we observe that they present more than one value for $g^{I}$, while
$g^{U}$ continues with a single value. We note that a set of values
of $g^{I}$ only appear for small values of $h$ and these tend to
decrease as $N$ increases. To observe the effects of these values
on the minimum of potentials, we consider an example where $N=10$,
$h=0.06$, $g_{1}^{I}=0.6771$, $g_{2}^{I}=0.3554$ and $g^{U}=0.3610$,
as we shown in figure\,\ref{fig:Plot of multipotentials}. We observe
that there is not much difference in the minimum of the effective
potential for the values of $g_{2}^{I}$ and $g^{U}$ considered in
this example. Finally, the plot in figure\,\,\ref{fig:Plot of multipotentials}
also exemplifies the fact that, for several of the solutions defined
by $g^{U}$ and $g^{I}$, the point $\sigma=\mu$ is actual a meta-stable
local minima, and not the global minima, which actually appears for
$0<\sigma<\mu$.

It is interesting to note the deep differences in the general properties
of the DSB mechanism, and quantitative aspects of the mechanism, in
the case of the improved effective potential. We also point out that
our results are in general compatible with the results obtained in
three and four dimensional space-time models, where the improved effective
potential was also calculated from the RGE, in the approximation of
leading logarithms\,\citep{Dias2010,Quinto2016,Dias2014,Souza2020,PhysRevD.72.037902,CHISHTIE2007}.

We close this section by pointing out the fact that common artifacts
of the perturbative calculation of the effective potential are non-convexity
and even instabilities (i.e., the potential not being bounded from
below). One notable case of this last problem is the so-called conformal
limit of the Standard Model, where the inclusion of the top quark
contribution to the one loop perturbative effective potential lead
to an unstable potential, a problem that was solved by summing up
the leading logs corrections using the RGE\,\citep{Elias2003}. Additional
improvements of this idea were further developed, and actually led
to a calculation of the Higgs mass of $M_{H}=141\text{GeV}$, not
far from the experimental value of $125\text{GeV}$\,\citep{Steele2013}.
We can also quote\,\citep{Meissner:2006zh,Meissner:2008uw} for showing
how an improved calculation of the effective potential may cure these
ailments.

\section{\label{sec:Conclusion}Conclusions}

In this paper we have studied the behavior of the unimproved and improved
effective potential in a massless $\left(1+1\right)$ dimensional
Gross-Neveu model with $N$ fermions fields. We have observed that
the improvement of the effective potential, which we calculated up
to the sixth power of the coupling constants, leads to different results
in comparison with the unimproved case. As a general rule, the use
of the RGE allows us to obtain higher order corrections to the effective
potential, based on the knowledge of the renormalization group functions
calculated up to some loop level (three in the case we considered
here \,\citep{Luperini1991}), and this could lead to a better understanding
of the DSB mechanism.

We notice that the improvement that we have performed in this work
has not been able to fully avoid such problems of the perturbative
effective potential. We can see in Fig.\,\ref{fig:Plot of multipotentials}
one of the improved potentials failing to be convex in the region
between two local minima. These potentials might also become unstable
for larger values of $\sigma$. We believe this comes from the fact
that we were able to sum up only contributions up to six loop order
in the $V_{\mathrm{eff}}^{I}\left(\sigma\right)$. A different summation
scheme, closer to the one adopted in\,\citep{Elias2003,Dias2010},
might allow for summing up infinite sub-series of higher loop order
contributions to $V_{\mathrm{eff}}^{I}\left(\sigma\right)$, and that
would probably eliminate at least some of these problems. This is
one topic we want to discuss in a future work.

Another interesting perspective is to incorporate a term associated
with the chemical potential: usually this appears as a mass parameter
associated with fermions, and it was not considered in the model studied
here, since the RGE improvement is simpler when the starting Lagrangian
is scale invariant. It has been reported in the literature that the
chemical potential is a key ingredient in the study of the polyacetylene
properties, corresponding for example to the doping concentration,
as discussed in\,\citep{Chodos1994,Caldas,Caldas2009,Caldas2009a,Kneur2007}
up to one loop order. Therefore, the idea would be to observe the
behavior of the effective potential when it has an explicit dependence
on the chemical potential at higher loop orders. This problem would
involve a multi-scale approach to the RGE improvement, as discussed,
for example, in\,\citep{Ford:1996hd,Chataignier:2018aud}. The presence
of a dimensional constant in the starting Lagrangian leads to the
appearance of two independent logarithms in the perturbative expression
for the effective potential since there would be, in general, contributions
involving also $\ln\left[\frac{m}{\mu}\right]$, with $m$ the fermion
mass, related to the chemical potential. That is another topic we
intent to investigate further.
\begin{acknowledgments}
The authors would like to thank André Lehum for his comments about
the manuscript, as well as the referee that provided very insightful
comments that helped us to improve our paper. This work was partially
supported by \emph{Fondo Nacional de Financiamiento para la Ciencia,
la Tecnología y la Innovación \textquotedbl Francisco José de Caldas\textquotedbl ,
Minciencias }Grand No. 848-2019\emph{ }(AGQ) and by \emph{Conselho
Nacional de Desenvolvimento Científico e Tecnológico} (CNPq), grant
305967/2020-7 (AFF).
\end{acknowledgments}

\bibliographystyle{unsrt}
\bibliography{Bibilografia}

\end{document}